\documentclass[final,5p,times,twocolumn]{elsarticle}
\usepackage{graphics}
\usepackage{amssymb}
\usepackage{color} 

\journal{Physica C}

\begin{document}

\begin{frontmatter}

\title{Bulk Superconductivity at 2.6 K in Undoped RbFe$_2$As$_2$}

\author[label1]{Z.~Bukowski}
\ead{bukowski@phys.ethz.ch}
\author[label2]{S.~Weyeneth}
\author[label3]{R.~Puzniak}
\author[label1]{J.~Karpinski}
\author[label1]{and B.~Batlogg}
\address[label1]{Laboratory for Solid State Physics, ETH Zurich, CH-8093 Zurich, Switzerland}
\address[label2]{Physik-Institut der Universit\"at Z\"urich, Winterthurerstrasse 190, CH-8057 Z\"urich, Switzerland}
\address[label3]{Institute of Physics, Polish Academy of Sciences, Aleja Lotnikow 32/46, PL-02-668 Warsaw, Poland}

\begin{abstract}
The iron arsenide RbFe$_2$As$_2$ with the ThCr$_2$Si$_2$-type structure is found to be a bulk superconductor with $T_{\rm c}=2.6$ K. The onset of diamagnetism was used to estimate the upper critical field $H_{\rm c2}(T)$, resulting in $\mu_0dH_{\rm c2}/dT\simeq-1.4$ T/K and an extrapolated $\mu_0H_{\rm c2}(0)\simeq2.5$ T. As a new representative of iron pnictide superconductors, superconducting RbFe$_2$As$_2$ contrasts with BaFe$_2$As$_2$, where the Fermi level is higher and a magnetic instability is observed. Thus, the solid solution series (Rb,Ba)Fe$_2$As$_2$ is a promising system to study the cross-over from superconductivity to magnetism.
\end{abstract}

\begin{keyword}
RbFe$_2$As$_2$ \sep iron pnictides \sep upper critical field \sep transition temperature \sep superconductivity
\PACS 74.70.Dd \sep 74.25.Op 

\end{keyword}

\end{frontmatter}

\section{Introduction}
The family of iron oxyarsenide $Ln$FeAsO$_{1-x}$F$_y$ ($Ln$ = Lanthanide element) exhibits  superconductivity with a maximum $T_{\rm c}$ up to 56 K \cite{2008_Kamihara_JACS_a, 2008_Ren_CPL_a}. Additionally, the iron-arsenide compounds $A$Fe$_2$As$_2$ ($A$ = alkaline earth element), crystallizing in the ThCr$_2$Si$_2$-type structure, are known to become superconducting with $T_{\rm c}$'s up to 38 K upon alkali metal substitution for the $A$ element \cite{2008_Rotter_PRL_a, 2008_Sasmal_PRL_a, 2009_Bukowski_PRB_a}, or partial transition metal substitution for Fe \cite{2008_Sefat_PRL_a}. In contrast to undoped BaFe$_2$As$_2$ with a magnetic ground state, superconductivity with relatively low $T_{\rm c}$'s was reported in the undoped alkali metal iron-arsenides KFe$_2$As$_2$ ($T_{\rm c}=3.8$ K) and CsFe$_2$As$_2$ ($T_{\rm c}=2.6$ K) \cite{2008_Sasmal_PRL_a}. Interestingly, RbFe$_2$As$_2$ is known to exist as well \cite{1984_Wenz_ZN_a}, although its physical properties have not been reported so far. Here we report on the superconductivity in undoped alkali metal iron arsenide RbFe$_2$As$_2$.
\begin{figure}[b!]
\centering
\vspace{-0cm}
\includegraphics[width=0.9\linewidth]{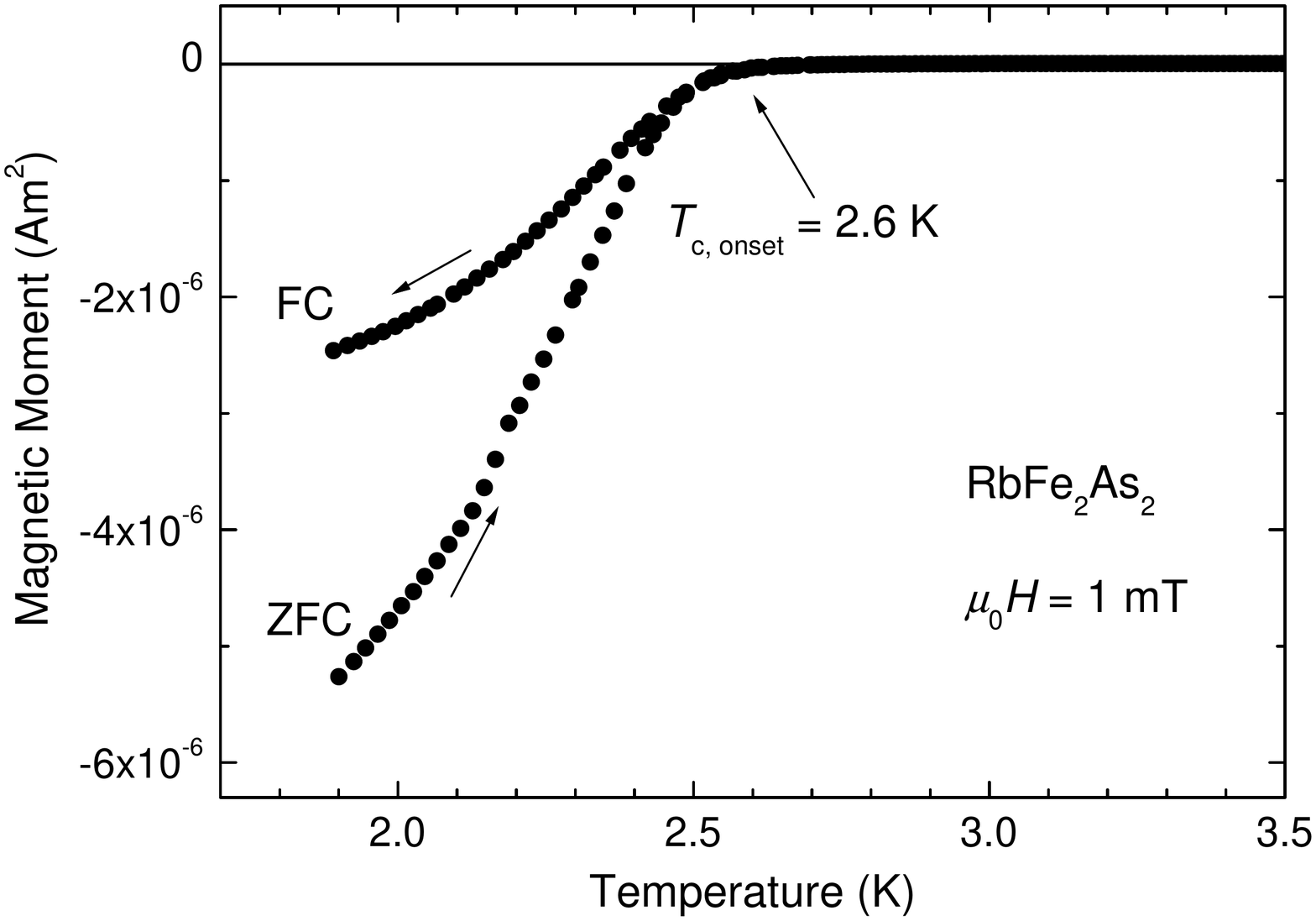}
\caption{Temperature dependence of the magnetic moment of a RbFe$_2$As$_2$ polycrystalline sample, measured in a magnetic field of 1 mT. Superconductivity sets in at $T_{\rm c}\simeq2.6$ K.
}
\label{fig1}
\end{figure}
\begin{figure}[t!]
\centering
\vspace{-0cm}
\includegraphics[width=0.9\linewidth]{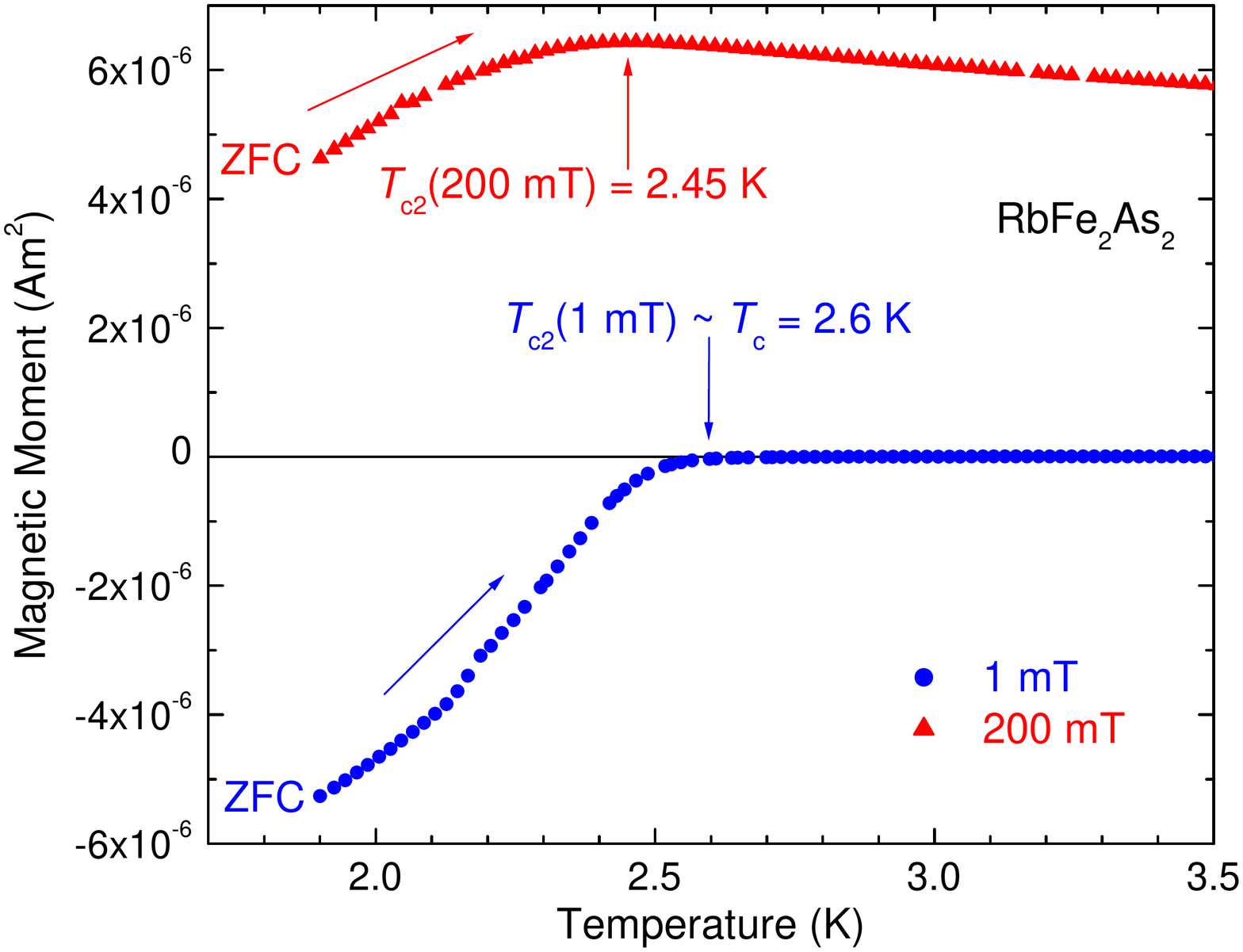}
\caption{Representative data of the magnetic moment used for the determination of $H_{\rm c2}(T)$, here for 1 mT and 200 mT, measured in the ZFC mode. A relative shift of the onset of superconductivity of 0.15 K is observed. An additional magnetic moment in the normal state in the 200 mT measurement, originates from a major normal state magnetic contribution.}
\label{fig1}
\end{figure}
\section{Experimental Details}
Polycrystalline samples of RbFe$_2$As$_2$ were synthesized in two steps. First, RbAs and Fe$_2$As were prepared from pure elements in evacuated and sealed silica tubes. Then, appropriate amounts of RbAs and Fe$_2$As were mixed, pressed into pellets and annealed at 650 $^\circ$C for several days in evacuated and sealed silica ampoules. Powder X-ray diffraction analysis revealed, that the synthesized RbFe$_2$As$_2$ is single phase material with lattice parameters $a=3.863$ \AA\ and $c=14.447$ \AA. Magnetization data have been recorded using a Quantum Design MPMS XL SQUID Magnetometer, equipped with a Reciprocating Sample Option.
\section{Results and Discussion} 
A polycrystalline sample of RbFe$_2$As$_2$ was studied for its low temperature magnetic properties. In Fig. 1 the magnetic moment in the field-cooled state (FC) and in the zero-field cooled state (ZFC) in a magnetic field of 1 mT are shown. The data are indicative of bulk superconductivity. The distinct onset of diamagnetism due to superconductivity is observed at $T_{\rm c}\simeq2.6$ K. Due to the limited temperature range of the equipment, the full development of the Meissner state could not be recorded. Nevertheless, the observed ZFC diamagnetic response mirrors bulk superconductivity and is consistent with the sample dimensions. The pronounced difference between the ZFC and FC curves stemms from remarkable flux-pinning in the sample, suggesting rather high critical current density.\\
The upper critical field $H_{\rm c2}$ was estimated from magnetization measurements performed at various magnetic fields in the mixed state. In Fig. 2, two representative measurements of the magnetic moment versus temperature are displayed for $\mu_0H=1$ mT and for $\mu_0H=200$ mT. We defined the upper critical field $H_{\rm c2}$ as the magnetic field $H$, where $T_{\rm c2}(H)$ is located. An obvious shift of the onset of superconductivity of 0.15 K is observed between the respective fields. In addition to the diamagnetic signal due to superconductivity, a distinct paramagnetic response develops due to the normal state magnetic contribution, rendering an accurate determination of $H_{\rm c2}(T)$ rather difficult. Nevertheless, since a clear downward curvature is observed due to the onset of superconducting diamagnetism, the trend of $H_{\rm c2}(T)$ can be followed down to 2 K. Figure 3 shows a summary of the results up to a field of 0.8 T, anticipating a linear slope close to $T_{\rm c}$ of $\mu_0dH_{\rm c2}/dT\simeq-1.4$ T/K. Assuming a simple WHH temperature dependence \cite{1966_Werthamer_PR_a}, which is known not to be applicable for the Fe pnictide superconductors with much higher transition temperatures, one would extrapolate $\mu_0H_{\rm c2}(0)\simeq2.5$ T, in comparision to the lower critical field $\mu_0H_{\rm c1}(0)\simeq4$ mT, as we estimated from field dependent initial magnetization curves, and the thermodynamic critical field $\mu_0H_{\rm c}(0)\simeq100$ mT. Superconductivity is, obviously, of type II.\\
The solid solution (Rb,Ba)Fe$_2$As$_2$ offers a particularly simple example where the interrelation between magnetic and superconducting ground states in the Fe pnictides can be studied through the controlled shift of the Fermi level. BaFe$_2$As$_2$ shows antiferromagnetic ordering competing with superconducting state. Appearently, doping of RbFe$_2$As$_2$ with Ba leads to a natural picture of enhancing $T_{\rm c}$ in the superconducting state, as the charge carrier concentration is varied. The appearence of superconductivity in RbFe$_2$As$_2$ opens up the window for a new interpretation of the occurence of superconducting state in (Rb,Ba)Fe$_2$As$_2$ \cite{2009_Bukowski_PRB_a, 2009_Karpinski_PHYSC_a}. 
\begin{figure}[t!]
\centering
\vspace{-0cm}
\includegraphics[width=0.9\linewidth]{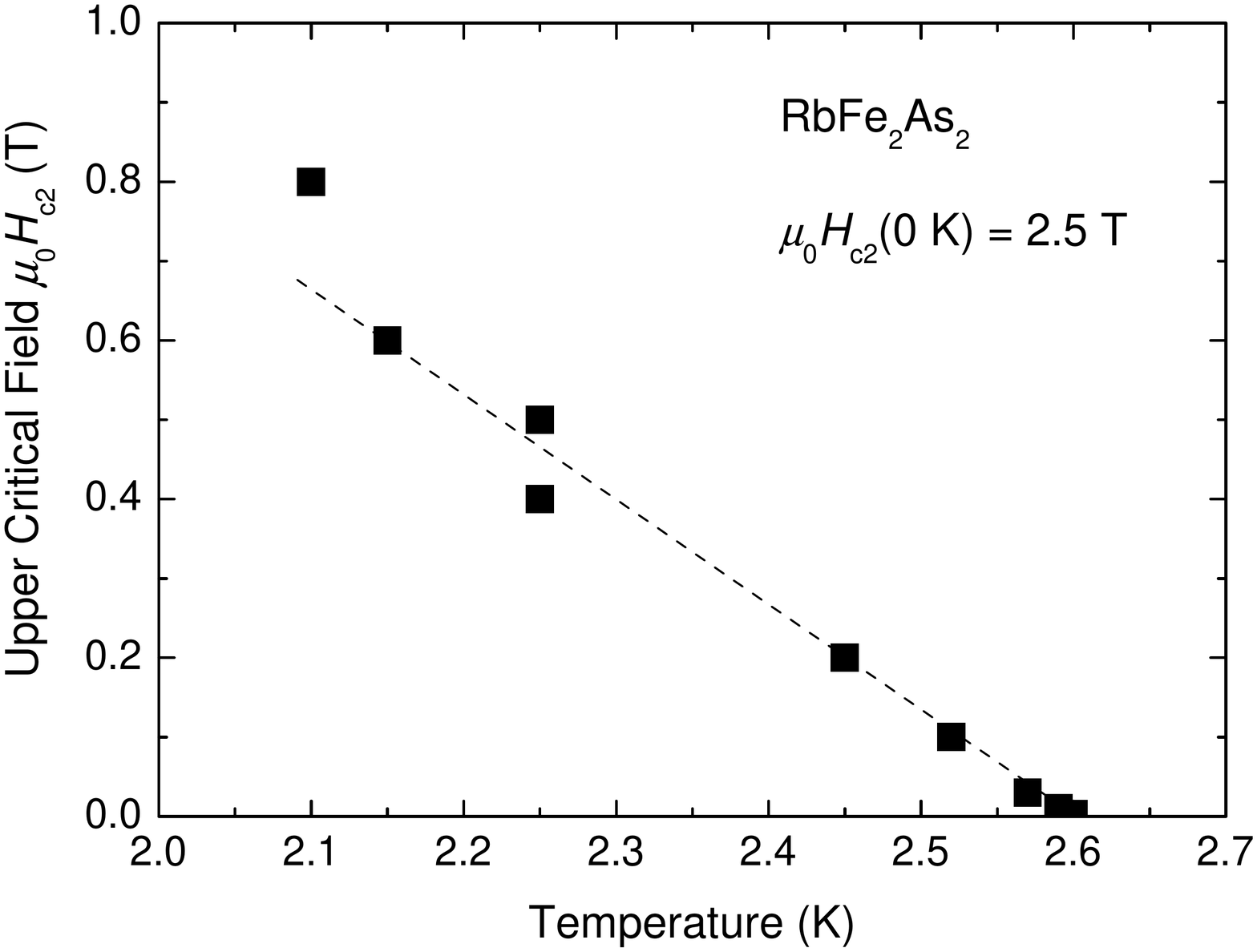}
\caption{Temperature dependence of $H_{\rm c2}$ for RbFe$_2$As$_2$. The estimate of $\mu_0H_{\rm c2}(0)\simeq2.5$ T is made using the WHH-approach.}
\label{fig1}
\end{figure}
\section{Conclusions}
Superconductivity is observed in undoped RbFe$_2$As$_2$ with a $T_{\rm c}\simeq2.6$ K. In this sense, it is useful to consider RbFe$_2$As$_2$ as a superconductor, located at the opposite end to the nonsuperconducting compound BaFe$_2$As$_2$ in the (Rb,Ba)Fe$_2$As$_2$ system. Therefore, superconductivity is enhanced by doping of an initially superconducting nonmagnetic parent compound. The upper critical field at zero temperature of RbFe$_2$As$_2$ is estimated to be $\mu_0H_{\rm c2}(0)\simeq2.5$ T.

\section{Acknowlegdements}
This work was supported by the Swiss National Science Foundation, by the NCCR program MaNEP, and partially by the Polish Ministry of Science and Higher Education within the research project for the years 2007-2009 (Grant No.
N N202 4132 33).

\end{document}